\documentclass[twocolumn,showpacs,aps,floatfix,superscriptaddress]{revtex4}
\usepackage{amsmath,amssymb,graphicx,eucal,bm}
\begin{document}
\title{Dynamics of Random Graphs with Bounded Degrees}
\author{E.~Ben-Naim}
\affiliation{Theoretical Division and Center for Nonlinear
Studies, Los Alamos National Laboratory, Los Alamos, New Mexico
87545, USA}
\author{P.~L.~Krapivsky}
\affiliation{Department of Physics,
Boston University, Boston, Massachusetts 02215, USA}
\begin{abstract}
We investigate the dynamic formation of regular random graphs. In our
model, we pick a pair of nodes at random and connect them with a link
if both of their degrees are smaller than $d$. Starting with a set of
isolated nodes, we repeat this linking step until a regular random
graph, where all nodes have degree $d$, forms.  We view this process
as a multivariate aggregation process, and formally solve the
evolution equations using the Hamilton-Jacobi formalism.  We calculate
the nontrivial percolation thresholds for the emergence of the giant
component when $d\geq 3$.  Also, we estimate the number of steps until
the giant component spans the entire system and the total number of
steps until the regular random graph forms.  These quantities are non
self-averaging, namely, they fluctuate from realization to realization
even in the thermodynamic limit.
\end{abstract}
\maketitle

\section{Introduction}             

A random graph is a set of nodes that are connected by random links
\cite{sr,er,bb,jlr}. When the number of links exceeds a certain
threshold, a giant component with a finite fraction of all nodes
emerges \cite{jklp,psw}. Therefore, random graphs are equivalent to a
mean-field percolation process \cite{sa,gg}.  Random graphs underlie
many natural phenomena from polymerization \cite{pjf,whs,pf} to the
spread of infectious diseases \cite{pg,mej,tz}, and they are also used
to model social networks \cite{ssw,nws,gn} as well as complexity of
algorithms \cite{bbckw}.

The classical random graph has no restrictions on the degree of a
node.  In many situations, the number of connections is limited, and
specifically, the degree of each node must be bounded.  Examples
include the percolating network of contacts in a bead pack \cite{hms},
computer networks \cite{dpp}, and communication networks \cite{wl}.
  
\begin{figure}[t]
\includegraphics[width=0.45\textwidth]{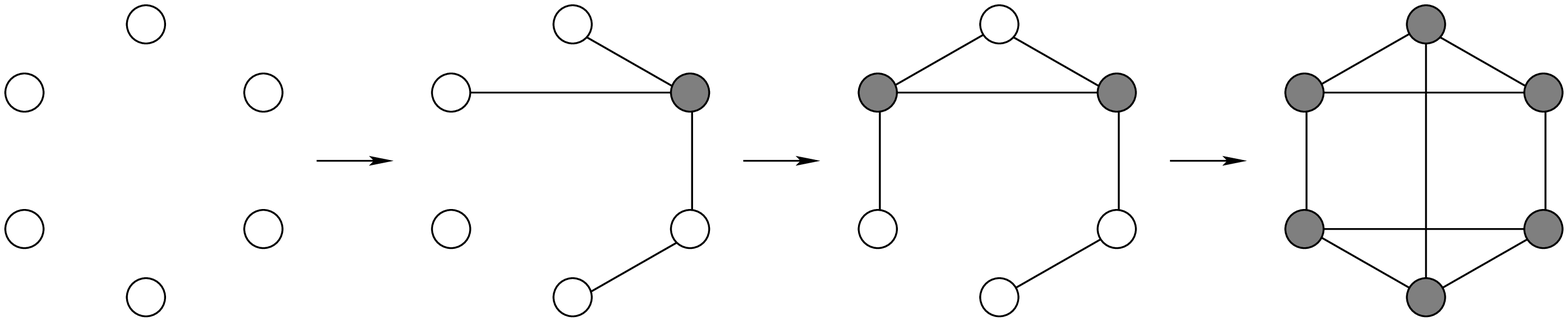}
\caption{Illustration of the linking process. Empty circles indicate
active nodes, and filled circles depict inactive nodes. The initial
state is a set of isolated nodes (left), while the final state is a
regular random graph (right). As links are added, connected components
form, and eventually, a single connected component spans the entire
system.}
\label{fig-process}
\end{figure}

We modify the canonical model for an evolving random graph
\cite{er,bb,jlr,bk05,aal,krb} by imposing a strict bound on the degree
of a node \cite{bq,rnw}. In an evolving random graph, two nodes are
picked at random, and regardless of their degrees, the nodes are
connected by a link. At the end of this process, a complete graph
forms.  In this study, we focus on a constrained process where the two
nodes are linked only if both degrees are smaller than $d$. Starting
with a set of $N$ isolated nodes, the final state is a regular random
graph \cite{ncw,rw}, where all nodes have degree $d$
(Fig.~\ref{fig-process}).  The evolving graph includes two types of
nodes, active nodes with degree smaller than the maximum, and inactive
nodes with the maximal degree. We obtain the density of active nodes
from the full degree distribution, which is shown to be a truncated
Poisson distribution.

In general, the graph is a set of disjoint connected components
\cite{gc}. In each component, any two nodes are connected by a path.
As in classical random graphs, our graphs could be in one of two
phases --- a non-percolating phase where all components are finite,
and a percolating phase where a single {\em giant} component with a
macroscopic number of nodes coexists with finite components. The
transition between these two phases occurs when the number of links
exceeds a threshold. For example, when $d=3$, the critical link
density, $L_g$, is
\begin{equation}
\label{Lg}
L_g=0.577200.
\end{equation}
When $d=1$ the final state consists of dimers, while when $d=2$ the
final state includes multiple rings \cite{bk11,bkun}. In both cases, 
the system does not condense into a single component.  We study the
more interesting situation \hbox{$d\geq 3$} where the system undergoes
a percolation transition and the final state is a fully connected
graph.

As a result of the linking process, connected components undergo an
aggregation process. In the traditional binary aggregation framework
\cite{mvs,sc}, connected components are characterized by the total
number of nodes \cite{bk05,aal,krb}.  In our process, however, this
minimal description does not lead to closed evolution equations.
Instead, we describe clusters by the total number of nodes of degree
$0$, $1$, $\ldots$, $d$.  This complete description naturally lends
itself to closed evolution equations.  We use the elegant
Hamilton-Jacobi method to obtain a formal solution for the density of
connected components, and find the percolation thresholds from this
solution.

We also use the degree distribution to illuminate the behavior during
the late stages of the evolution. The regular random graph forms via
a two-step process. First, the giant component takes over the entire
system. This step occurs after $T$ linking attempts, with $T\sim
N/(\ln N)^{d-1}$.  At this point, only a small minority of the nodes
in the giant component remain active. The total number of active
nodes, $M$, scales logarithmically with system size,
\begin{equation}
\label{M-sol}
M\sim (\ln N)^{d-1}. 
\end{equation}
Eventually all nodes become inactive, and the regular random graph is
complete. The number of linking attempts required to reach this final
state is proportional to the system size. Hence, the evolution of the
regular random graph is much slower than the evolution of the
classical random graph, where $T$ is logarithmic in $N$ \cite{krb}.

The rest of this paper is organized as follows. The degree
distribution is derived in section \ref{DD}. We then show that the
linking process is equivalent to a multivariate aggregation process
(Sec.~\ref{MA}). A formal solution is obtained in Sec.~\ref{HJ} using
the Hamilton-Jacobi method. The emergence of the giant component is
discussed in section \ref{GC}.  In particular, the mass of the giant
component is obtained as a function of time. The late-time kinetics
including the emergence of a fully connected graph and a perfect
regular random graph are described in section \ref{LT}. We discuss the
results in section \ref{disc}. In Appendix A, we rederive relevant
properties of classical random graphs using the Hamilton-Jacobi
method.

\section{Degree Distribution}
\label{DD}

We study a random process that generates a regular random graph. The
process starts at time $t=0$ with $N$ disconnected nodes. At each
elementary step, two nodes are chosen at random. If the degrees of the
two nodes are both smaller than $d$, a new link connects them;
otherwise, no link is added.  Time is augmented by $2/N$ after each
linking attempt, \hbox{$t\to t+2/N$}, so that every node participates
in one linking attempt per unit time, on average. Linking is
repeatedly attempted until a regular random graph forms
(Fig.~\ref{fig-process}).

There are two types of nodes --- active nodes with degree smaller than
$d$ and inactive nodes with degree equal to $d$. A new link between
two active nodes of degrees $i<d$ and $j<d$ augments the degrees:
\begin{equation}
\label{linking}
(i,j)\to (i+1,j+1)\,.
\end{equation}
This random process occurs with unit rate.  The linking process
\eqref{linking} transforms the system from a set of active nodes into
a set of inactive nodes, as illustrated on Fig.~\ref{fig-process}.

Let $n_j(t)$ be the degree distribution, that is, the fraction of
nodes with degree $j$ at time $t$.  The degree distribution is
properly normalized, $\sum_{j=0}^d n_j=1$, and its partial sum gives
the total density of active nodes, $\nu$,
\begin{equation}
\label{nu-def}
\nu=\sum_{j=0}^{d-1} n_j.
\end{equation}
The density of active nodes controls the process: the degree
distribution obeys the rate equations
\begin{equation}
\label{nj-eq}
\frac{d n_j}{dt}=
\begin{cases}
\nu\, (n_{j-1}-n_j)    \qquad& j<d,\\
\nu\, n_{d-1}            \qquad& j=d.
\end{cases}
\end{equation}
The initial condition is 
\begin{equation}
\label{n_initial}
n_j(0)=\delta_{j,0}\,,
\end{equation}
and the ``boundary'' condition $n_{-1}\equiv 0$ ensures that
\hbox{$dn_0/dt = - \nu\,n_0$}.  Equations \eqref{nj-eq} are nonlinear
because \eqref{linking} is a two-body process. Yet, since the density
$\nu$ merely sets the overall linking rate, we can linearize
Eqs.~\eqref{nj-eq} by introducing the time variable
\begin{equation}
\label{tau-def}
\tau=\int_0^t dt'\, \nu(t'),
\end{equation}
or equivalently, $d\tau/dt=\nu$.  In terms of the time variable $\tau$, the
rate equations become
\begin{equation}
\label{nj-lin}
\frac{dn_j}{d\tau}=
\begin{cases}
n_{j-1}-n_j\qquad& j<d,\\
n_{d-1}    \qquad& j=d.
\end{cases}
\end{equation}

We use the initial condition \eqref{n_initial} and solve
\eqref{nj-lin} recursively to find $n_0=e^{-\tau}$,
$n_1=\tau\,e^{-\tau}$, \hbox{$n_2=\tfrac{1}{2}\tau^2e^{-\tau}$}, etc.
Therefore, the degree distribution is a truncated Poisson distribution
\cite{nw}
\begin{equation}
\label{nj-sol}
n_j=\frac{\tau^j}{j!}e^{-\tau},
\end{equation}
for $j<d$.  The density of inactive nodes, \hbox{$n_d=1-\nu$}, follows
from the normalization condition. From the definitions \eqref{nu-def}
and \eqref{tau-def}, the time variable $\tau$ obeys the nonlinear
ordinary differential equation
\begin{equation}
\label{tau-eq}
\frac{d\tau}{dt} =\sum_{j=0}^{d-1} \frac{\tau^j}{j!} e^{-\tau},
\end{equation}
with the initial condition $\tau(0)=0$.  Equations \eqref{nj-sol} and
\eqref{tau-eq} specify the degree distribution.

\begin{figure}[t]
\includegraphics[width=0.45\textwidth]{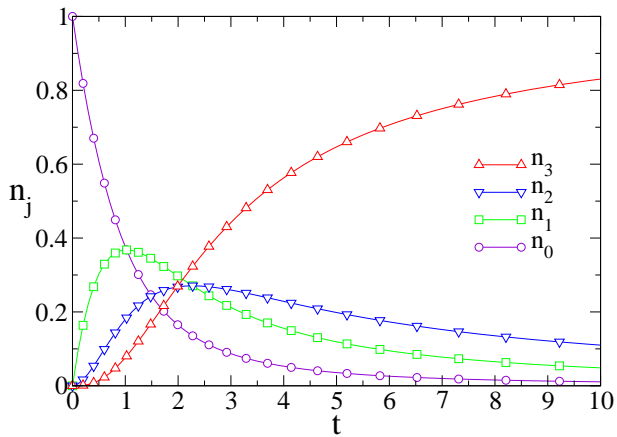}
\caption{The degree distribution $n_j(t)$ versus time $t$ for $d=3$.}
\label{fig-nt}
\end{figure}

The time evolution of the degree distribution, obtained by numerical
integration of \eqref{tau-eq} for the case $d=3$, is shown in
Fig.~\ref{fig-nt}. As expected, the density of isolated nodes,
$n_0(t)$, declines monotonically, while the density of inactive nodes,
$n_3(t)$, increases monotonically. Isolated nodes initially dominate,
then nodes with degree one are most numerous, and eventually, inactive
nodes take the lead for good.

The average degree, $\langle j\rangle =\sum_{j=0}^d j n_j$,
characterizes the overall connectivity.  Since every link connects two
nodes, the density of links, $L$, is proportional to the average
degree, $L=\langle j\rangle/2$. By multiplying \eqref{nj-lin} by $j$
and summing over $j$, we see that the average degree satisfies
\begin{equation}
\label{jav-eq}
\frac{d\langle j\rangle}{d\tau} = \nu .
\end{equation}
Using the initial and final values, $\langle j\rangle_0=0$ and
$\langle j\rangle_\infty =d$, we have the integral identity
$\int_0^\infty d\tau\, \nu(\tau)=d$.

\section{Multivariate Aggregation}
\label{MA}

Every node belongs to a connected component, or ``cluster''.
Initially, there are $N$ clusters, and eventually, a single cluster
remains. The number of clusters dwindles as a result of a
cluster-cluster aggregation \cite{fl} process: any link between two
active nodes in separate clusters merges the two clusters.  Since the
aggregation rate depends on the total number of {\em active} nodes,
characterization of clusters by their total size is insufficient.
Adding the number of active nodes as a second variable does not lead
to a closed description either, because the number of active nodes may
change once a link is added.  Therefore, we use a complete description
and characterize clusters by $(d+1)-$dimensional vectors
\begin{equation}
{\bm k} \equiv (k_0,k_1,\cdots,k_d),
\end{equation}
where $k_j$ is the number of nodes of degree $j$.  The cluster size,
$k$, is simply \hbox{$k \equiv |{\bm k}| = k_0 + k_1 +\cdots +
k_d$}. Finite clusters are typically trees \cite{tree}. Since there
are \hbox{$k-1$} links in a tree of size $k$, we have the constraint
\begin{equation*}
2(k-1)=k_1+2k_2+\ldots+dk_d.
\end{equation*}
In particular, minimal clusters with \hbox{${\bm k}=(1,0,\cdots,0)$}
trivially satisfy this condition.

As a result of the linking process \eqref{linking}, clusters undergo
the multivariate aggregation process 
\begin{equation}
\label{aggregation}
{\bm l},{\bm m}\ {\buildrel l_i m_j\over \longrightarrow}\
{\bm l}+{\bm m}+{\bm e}_i+{\bm e}_j,
\end{equation}
with $i<d$ and $j<d$.  Here, the vector ${\bm e}_j$ describes the
degree augmentation $j\to j+1$,
\begin{eqnarray*}
{\bm e}_0       &=& (-1,1,0,0,\ldots,0),\\
{\bm e}_1       &=& (0,-1,1,0,\ldots,0),\\
&\vdots&\\
{\bm e}_{d-1} &=& (0,0,\ldots,0,-1,1).
\end{eqnarray*}
The aggregation rate $l_i\times m_j$ in \eqref{aggregation} equals the
product of the number of nodes of degree $i$ in the first cluster and
the number of nodes of degree $j$ in the second cluster.

The density $c({\bm k},t)$ of clusters with ``size'' ${\bm k}$ at time
$t$ obeys the rate equation
\begin{eqnarray*}
\frac{d c({\bm k})}{dt} \!=\!
\frac{1}{2}\!\!\sum_{\substack{{\bm l},{\bm m}\\i,j<d}}\!\!
l_i m_j
c({\bm l})c({\bm m})
\delta_{{\bm k},{\bm l}+{\bm m}+ {\bm e}_i+{\bm e}_j} \!-
\nu\!\left(\sum_{j=0}^{d-1} k_j\!\!\right)\!  c({\bm k}),
\end{eqnarray*}
where $\nu$ is the density of active nodes \eqref{nu-def}. Henceforth,
the time dependence is implicit. The gain term directly reflects the
aggregation process \eqref{aggregation}, and the factor $1/2$ prevents
over-counting.  The loss rate is proportional to the total number of
active nodes in the cluster, as well as the density of active nodes in
the entire system. 

A crucial feature of the above rate equation is that the gain term and
the loss term are different in nature. The gain term, which describes
creation of clusters with finite size, is proportional to the product
of cluster densities. The loss term, which describes loss of finite
clusters due to linking, is proportional to the product of the cluster
density and the total density of active nodes in the entire system,
which does take into account the giant component, if one exists.  For
this reason, the rate equation is valid in the non-percolating phase
as well as in the percolating phase.

In terms of the time variable $\tau$, the loss term becomes linear
\begin{equation}
\label{ck-eq}
\frac{d c({\bm k})}{d\tau} \!=\!  \frac{1}{2\nu}\!\!\sum_{\substack{{\bm
l},{\bm m}\\i,j<d}}\!\!  l_i m_j c({\bm l})c({\bm m}) \delta_{{\bm
k},{\bm l}+{\bm m}+ {\bm e}_i+{\bm e}_j} \!-
\left(\sum_{j=0}^{d-1} k_j\!\!\right)\!  c({\bm k}). 
\end{equation}
The initial condition is $c({\bm k},0)=\delta_{{\bm
k},(1,0,\ldots,0)}$.  The rate equations for the cluster size
distribution are nonlinear, in contrast with the linear equations
\eqref{nj-lin} for the degree distribution. Moreover, the rate
equations \eqref{ck-eq} are substantially more challenging than those
previously analyzed in studies of multivariate aggregation processes
\cite{kb,mm}.

The generating function approach is generally useful in aggregation 
problems \cite{krb,fl,zem,bk05a}, and in our case, we must use the 
multivariate generating function
\begin{equation}
\label{cx-def}
C({\bm x},\tau)= \sum_{\bm k} c({\bm k},\tau)\,{\bm x}^{{\bm k}}.
\end{equation}
Here ${\bm x} \equiv (x_0,x_1,\cdots,x_d)$ and ${\bm x}^{{\bm
k}}\equiv x_0^{k_0} x_1^{k_1}\cdots x_d^{k_d}$ is a shorthand for the
product of monomials.  Multiplying \eqref{ck-eq} by ${\bm x}^{{\bm
k}}$ and performing the summation over ${\bm k}$, we find that the
generating function evolves according to
\begin{equation}
\label{cx-eq}
\frac{\partial C}{\partial \tau}=
\frac{1}{2\nu}\left(\sum_{j=0}^{d-1}x_{j+1}\,\frac{\partial
C}{\partial x_j}\right)^{\!2}
-\sum_{j=0}^{d-1}x_j\frac{\partial C}{\partial x_j}.
\end{equation}
The initial condition is 
\begin{equation}
\label{Cx_0}
C({\bm x}, 0) = x_0\,.
\end{equation}

\section{Hamilton-Jacobi Formulation}
\label{HJ}

The generating function satisfies a non-linear Partial Differential
Equation (PDE) with non-constant coefficients.  Crucially, equation
\eqref{cx-eq} is a {\em first-order} PDE, and hence, it can be reduced
to a set of coupled Ordinary Differential Equations (ODEs) by applying
the method of characteristics \cite{via,jdl}. This is a significant
simplification as a PDE is essentially a collection of infinitely many
ODEs. Moreover, in the present case, we conveniently treat the
governing equation as a {\em Hamilton-Jacobi} equation, thereby
bypassing the method of characteristics \cite{via,img}.  Then, the
canonical Hamilton equations coincide with the equations for the
characteristics.

First, we convert \eqref{cx-eq} into a Hamilton-Jacobi equation
\begin{equation}
\label{hj-eq}
\frac{\partial C({\bm x},\tau)}{\partial \tau} + 
H({\bm x}, \nabla C,\tau) = 0
\end{equation}
and view $C({\bm x},\tau)$ and $H({\bm x}, \nabla C,\tau)$ as
``action'' and ``Hamiltonian'', respectively. Second, we treat ${\bm
x}$ as ``coordinate'', and $\nabla C$ as ``momentum'', \hbox{${\bm p}
= \nabla C$}; in components, $p_j = \frac{\partial C}{\partial x_j}$
for all $j$.  The Hamiltonian is
\begin{equation}
\label{H-def}
H({\bm x}, {\bm p},\tau) = \sum_{j=0}^{d-1}x_jp_j 
- \frac{\Pi_1^2}{2\nu(\tau)}\,,
\end{equation}
where for convenience, we introduced a set of auxiliary functions,
\begin{eqnarray}
\label{Pi-def}
\Pi_j = \sum_{i=j}^{d}x_i\,p_{i-j},
\end{eqnarray}
for all $j$.  We stress that the Hamiltonian \eqref{H-def} is not a
conserved quantity because of the explicit dependence on time $\tau$,
which enters via $\nu(\tau)$. The Hamilton-Jacobi equation
\eqref{hj-eq} is equivalent to the canonical Hamilton equations
\begin{equation}
\label{can_H}
\frac{d x_j}{d\tau} = \frac{\partial H}{\partial p_j}\,, \qquad
\frac{d p_j}{d\tau} = -\frac{\partial H}{\partial x_j}\,.
\end{equation}
Hence, the non-linear partial differential equation \eqref{cx-eq}
reduces to $2(d+1)$ coupled ordinary differential equations for the
coordinates $x_j$ and the momenta $p_j$. 

From \eqref{can_H} and \eqref{H-def}, the evolution equations for the
momenta are
\begin{equation}
\label{pj-eq}
\frac{dp_j}{d\tau}=
\begin{cases}
\frac{\Pi_1}{\nu}\, p_{j-1} -p_j \qquad& j<d, \\
\frac{\Pi_1}{\nu}\, p_{d-1} \qquad& j=d.
\end{cases}
\end{equation}
The initial condition $p_j(0)=\delta_{j,0}$ follows from ${\bm p} =
\nabla C$ and \eqref{Cx_0}, and the boundary condition is $p_{-1}\equiv
0$.  The structure of the {\em linear} equations \eqref{pj-eq} for
$p_j$ is identical to that of equations \eqref{nj-lin} for $n_j$, and
the initial conditions are the same, too.  By integrating equations
\eqref{pj-eq} recursively, we find
\begin{equation}
\label{pj-sol}
p_j = \frac{u^j}{j!}\,e^{-\tau}
\end{equation}
for $j<d$, with function $u$ defined by
\begin{equation}
\label{u-def}
u=\int_0^\tau d\tau'\, \frac{\Pi_1(\tau')}{\nu(\tau')}\,,
\end{equation}
or alternatively, $du/d\tau=\Pi_1/\nu$.  To determine $p_d$, we
rewrite the last equation in \eqref{pj-eq} as $dp_d/du = p_{d-1}$, and
therefore, $p_d=\int_0^u du' \, p_{d-1}(u')$. Formally, the momenta
$p_j$ resemble the degree distribution $n_j$.

Using \eqref{can_H} and \eqref{H-def}, the evolution equations for the
coordinates are
\begin{equation}
\label{xj-eq}
\frac{dx_j}{d\tau}=
\begin{cases}
x_j - \frac{\Pi_1}{\nu}\, x_{j+1} \qquad& j<d,\\
0                                           \qquad & j=d.
\end{cases}
\end{equation}
The initial condition is ${\bm x}(\tau=0)={\bm y}$.  One of the
coordinates does not change, $x_d = y_d$. We may integrate
\eqref{xj-eq} and solve for the final coordinate ${\bm x}$ as a
function of the initial coordinate ${\bm y}$.  Yet, as shown below, we
need to solve the inverse problem --- find ${\bm y}$ as a function of
${\bm x}$.

Fortunately, the auxiliary functions defined in \eqref{Pi-def} enable
us to accomplish this task. We first note that at time $\tau=0$, these
sums coincide with the initial coordinates, $\Pi_j(\tau=0)=y_j$. Using
the evolution equations \eqref{pj-eq} and \eqref{xj-eq}, we calculate
the evolution equations for $\Pi_j$,
\begin{eqnarray*}
\dot \Pi_j &=&
\sum_{i=j}^{d} \left(\dot x_i p_{i-j} + x_{i}\dot p_{i-j}\right)\\
&=&\sum_{i=j}^{d-1} \left( x_{i}-\tfrac{\Pi_1}{\nu} x_{i+1}\right) p_{i-j}
+\sum_{i=j}^{d} x_{i}\left(\tfrac{\Pi_1}{\nu} p_{i-j-1}-p_{i-j}\right)\\
&=&-x_d\, p_{d-j},
\end{eqnarray*}
when $j>0$.  Here, the overdot is shorthand for time derivative,
$\dot{} \equiv \frac{d}{d\tau}$.  A similar calculation shows that
$\Pi_0$ is conserved, $\dot \Pi_0=0$.  Hence, the functions $\Pi_j$
obey
\begin{equation}
\label{Pij-eq}
\frac{d\Pi_j}{d\tau}=
\begin{cases}
0\qquad & j=0, \\
-x_d \, p_{d-j}\qquad & j>0.
\end{cases}
\end{equation}
The initial condition is $\Pi_j(0)=y_j$.  Since the coordinate $x_d$
is a constant, integration of these evolution equations is immediate,
and using the initial condition \hbox{$\Pi_j(0)=y_j$}, we find
\cite{back}
\begin{equation}
\label{yj-sol}
y_j=
\begin{cases}
\textstyle{\sum_{i=0}^{d} x_i p_i}\quad & j=0,\\
\textstyle{\sum_{i=j}^{d} x_{i} p_{i-j}} + 
x_d \int_0^\tau d\tau'  p_{d-j}(\tau')\quad & j>0.
\end{cases}
\end{equation}
(The lower expression holds up to $j=d$; indeed, using $p_0=e^{-\tau}$
we recover $y_d=x_d$.)  Equations \eqref{yj-sol}, together with the
known momenta \eqref{pj-sol}, specify the initial coordinates in terms
of the final coordinates. Remarkably, we formally solved the inverse
problem!

To complete the solution of the Hamilton-Jacobi equations, we must
specify the function $u$ in \eqref{pj-sol}. Using the definition
\eqref{u-def}, we calculate the second derivative of $u$,
\begin{equation*}
\frac{d^2u}{d\tau^2}=\frac{d}{d\tau}\frac{\Pi_1}{\nu}=
-x_d\frac{p_{d-1}}{\nu}+\frac{n_{d-1}}{\nu}\frac{du}{d\tau}.
\end{equation*}
Here, we used $d\nu/d\tau=-n_{d-1}$, as follows from
\eqref{nj-lin}. By multiplying the above equation by $\nu$, and by
substituting \eqref{nj-sol} and \eqref{pj-sol}, we find that $u$
satisfies a {\em closed} second-order ordinary differential equation,
\begin{equation}
\label{u-eq}
\left(\sum_{j=0}^{d-1} \frac{\tau^j}{j!}\right)
\frac{d^2u}{d\tau^2} -
\frac{\tau^{d-1}}{(d-1)!} \frac{du}{d\tau}
+
x_d\frac{u^{d-1}}{(d-1)!}=0.
\end{equation}
The initial conditions are $u(0)=0$ and $du(0)/d\tau=y_1$. This
equation is linear if and only if $d\leq 2$.

However, we still do not know $y_1$ because the initial coordinates
are coupled to the function $u$, according to \eqref{yj-sol}. To find
$y_1$, we have to integrate equation \eqref{u-eq}, starting with trial
initial conditions $u(0)=0$ and \hbox{$du(0)/d\tau=y_1^*$}, and find
the root of the integral equation
\begin{equation}
\label{root}
y_1^*=\sum_{i=1}^d x_ip_{i-1}+x_d\int_0^\tau d\tau'\, p_{d-1}(\tau').
\end{equation}
The second order ODE \eqref{u-eq} and the integral equation
\eqref{root} specify the variable $u$ and hence, the momenta $p_j$ 
and the initial coordinates $y_j$ via the explicit solutions
\eqref{pj-sol} and \eqref{yj-sol}. We stress that the function
$u\equiv u({\bm x},\tau)$ depends on the coordinate ${\bm x}$.

Finally, we obtain the generating function \eqref{cx-def} by
evaluating its complete derivative with respect to $\tau$,
\begin{eqnarray}
\label{cxt-eq}
\frac{dC({\bm x},\tau)}{d\tau} &=& \frac{\partial C}{\partial \tau} +
\frac{\partial C}{\partial {\bm x}} \cdot \frac{d {\bm x}}{d\tau} =- H +
{\bm p}\cdot \frac{\partial H}{\partial {\bm p}}\nonumber \\
&=&- \frac{\Pi_1^2}{2\nu} = -\frac{\nu}{2}\left(\frac{du}{d\tau}\right)^2.
\end{eqnarray}
By integrating \eqref{cxt-eq} subject to the initial condition $C({\bm
x},0)=y_0$, and recalling the conservation law \hbox{$y_0=\Pi_0$}, we
get
\begin{equation}
\label{cx-sol}
C ({\bm x},\tau)= \sum_{j=0}^d x_jp_j  -
\frac{1}{2}\int_0^\tau d\tau'\,\nu(\tau')\,\left(\frac{du}{d\tau}\right)^2.
\end{equation}
This expression, supplemented by equations \eqref{u-eq} and
\eqref{root} for $u$, as well as the explicit expressions
\eqref{pj-sol} for the momenta, constitutes a formal solution for the
generating function. Therefore, representation of the linking process
as a multivariate aggregation process enables analytical treatment of
the cluster-size distribution.

\section{Giant Component}
\label{GC}

By definition, the size distribution $c({\bm k})$ describes finite
connected components.  However, the system may be in one of two phases
--- a {\em non-percolating} phase in which finite components contain
all the mass, and a {\em percolating} phase in which a giant
component, containing a fraction $g$ of the entire mass, coexists with
finite components. We expect that the giant component emerges
suddenly, at a finite time $t_g$ (see Appendix A).

The mass of the giant component $g$ follows directly from the momenta
when ${\bm x}=(1,1,\ldots,1)$. Indeed, using \eqref{cx-def}, we have
\begin{equation}
\label{g-def}
1-g=\sum_{\bm k} k\, c({\bm k})=\sum_{j=0}^d p_j,
\end{equation}
or alternatively, $g=1-y_0$, as follows from the conservation law
$\Pi_0=y_0$. Hence, we substitute $x_d=1$ into equation \eqref{u-eq},
and obtain the evolution equation for $u$,
\begin{equation}
\label{u-eq-2}
\left(\sum_{j=0}^{d-1} \frac{\tau^j}{j!}\right)\frac{d^2u}{d\tau^2}
- \frac{\tau^{d-1}}{(d-1)!} \frac{du}{d\tau} + \frac{u^{d-1}}{(d-1)!}=0.
\end{equation}
The initial conditions are $u(0)=1$ and $du(0)/d\tau=y_1$. Also, when
${\bm x}=(1,1,\ldots,1)$, the integral equation \eqref{root} becomes
\begin{equation}
\label{root-2}
y_1^*=\sum_{j=0}^{d-1} p_j+\int_0^\tau d\tau'\, p_{d-1}(\tau'),
\end{equation}
with $p_j$ given by \eqref{pj-sol}.

First, there is the trivial solution $u=\tau$ and $y_j=1$ for all $j$,
for which the momenta equals the degree distribution, $p_j=n_j$. For
this solution, $g=0$, and the system is in the finite component
phase. When $d\leq 2$, the trivial solution holds at all times.  In
the case $d=1$ the final state is merely a collection of $N/2$ dimers;
in the case $d=2$, the final state includes only rings
\cite{bk11,bkun}.  In both of these cases the system does not condense
into a single component.

When $d\geq 3$, there is a second nontrivial solution when
$\tau>\tau_g$ or equivalently, $t>t_g$.  To find the nontrivial
solution, we solved equation \eqref{u-eq-2} numerically, using Adams'
method \cite{ba,dz}, starting with a trial initial condition $u(0)=0$
and $du(0)/d\tau=y_1^*$, and then found the root of the integral
equation \eqref{root-2} using the bisection method
\cite{ptvf}. (Conveniently, the integral equations may be converted
into first-order differential equations \cite{back}.)  We then
obtained the mass of the giant component using \eqref{g-def}. To
evaluate time, $t$, and the average degree, $\langle j\rangle$, in
terms of the modified time variable $\tau$, we simply integrated
$dt/d\tau=1/\nu$ and equation \eqref{jav-eq}.

For the case $d=3$, we find the critical time \hbox{$\tau_g=1.197395$}
that corresponds to
\begin{equation}
t_g=1.243785.
\end{equation}
When the giant component emerges, the average degree of a node is
$\langle j\rangle_g=1.154399$. Moreover, the critical density of links
quoted in \eqref{Lg} is obtained from $L_g=\langle j\rangle_g/2$.
Thus, the fraction $f_g$ of successful linking attempts is
\begin{equation}
f_g=\frac{\langle j \rangle_g}{t_g}=0.928135.
\end{equation}
Generally, the percolation parameters characterizing the phase
transition from the finite component phase to the giant component
phase are nontrivial.

Table I lists the percolation parameters for \hbox{$3\leq d\leq 7$}.
Remarkably, the restriction on the degree of a node has a small effect
on the percolation parameters already when $d=4$. Moreover, as the
maximal degree $d$ increases, the percolation parameters quickly
converge to the classical random graph values.

\begin{figure}[t]
\includegraphics[width=0.45\textwidth]{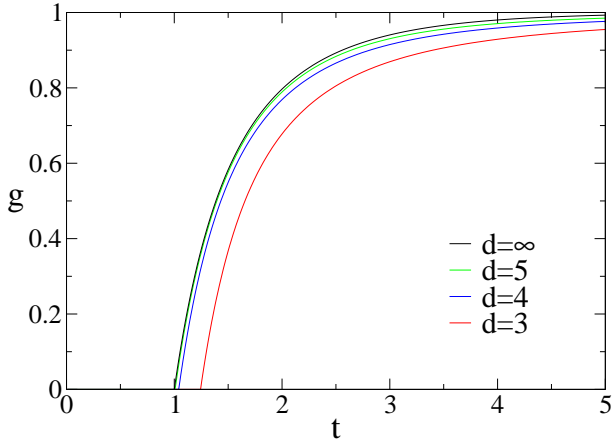}
\caption{The giant component mass $g(t)$ versus time $t$. Results of
  the Hamilton-Jacobi theory are shown for $d=3$, $4$, $5$. Also shown
  as reference, is the solution to Eq.~\eqref{rg-g-sol} for the
  classical random graph, $d=\infty$}
\label{fig-gt}
\end{figure}

Figure \ref{fig-gt} shows the mass of the giant component for
\hbox{$3\leq d\leq 5$} and $d=\infty$. At moderate times, the quantity
$g$ converges to the classical random graph behavior
\eqref{rg-g-sol}. Yet, as will be shown in the next section, the
long-time evolution of the regular random graph is much slower
compared with the classical random graph.

\begin{table}[t]
\begin{tabular}{|l|l|l|l|l|l|}
\hline
$d$ & $t_g$ & $\tau_g$ & $\langle j\rangle_g$ & $f_g$ & $\nu_g$  \\
\hline
3 & 1.243785 & 1.197395& 1.154399 & 0.928135 & 0.880052\\
4 & 1.041130 & 1.035995& 1.030922 & 0.990195 & 0.978725\\
5 & 1.007307 & 1.006593& 1.005879 & 0.998582 & 0.996238\\
6 & 1.001169 & 1.001074& 1.000978 & 0.999809 & 0.999403\\
7 & 1.000164 & 1.000152& 1.000141 & 0.999977 & 0.999917\\
$\infty$ & 1 & 1 & 1 & 1 & 1 \\
\hline
\end{tabular}
\caption{Percolation parameters for $3\leq d\leq 7$ and
\hbox{$d=\infty$}. Listed are the time, $t_g$, the modified time, 
$\tau_g$, the average degree, $\langle j\rangle_g=2L_g$, the fraction
of successful linking attempts, $f_g$, and the density of active nodes, 
$\nu_g$, when the giant component emerges.}
\end{table}

\begin{figure}[t]
\includegraphics[width=0.45\textwidth]{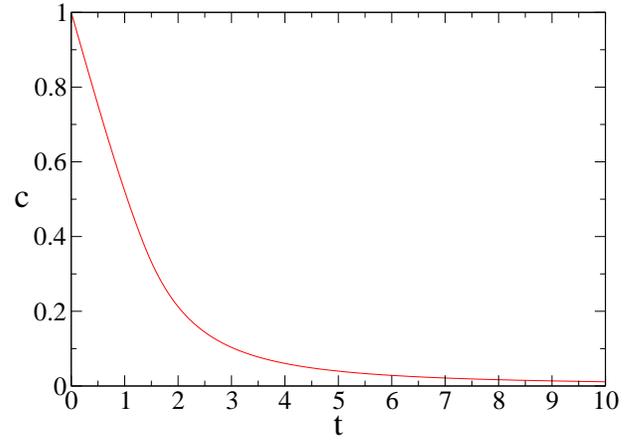}
\caption{The concentration $c(t)$ versus time $t$ for the case $d=3$.}
\label{fig-ct}
\end{figure}

The function $u$ at ${\bm x}=(1,1,\ldots,1)$ also gives the total
density of clusters, $c$, since \hbox{$c= C(1,1,\ldots,1)$}.  The
formal solution \eqref{cx-sol} together with \eqref{g-def} give
\begin{equation}
\label{ct}
c(\tau)= 1- g(\tau) - 
\frac{1}{2}\int_0^\tau d\tau'\,\nu(\tau')\,\left(\frac{du}{d\tau}\right)^2.
\end{equation}
Figure \ref{fig-ct} shows that the concentration is a smooth function
of time, even at the percolation point \cite{smooth}. We also mention
a useful relation between the cluster density and the link density
throughout the non-percolating phase.  Every link reduces the number
of clusters by one. Hence, $dc/dt=-dL/dt$, and consequently, $c+L=1$
when $t\leq t_g$. In particular, $c_g+L_g=1$.

\begin{figure}[t]
\includegraphics[width=0.45\textwidth]{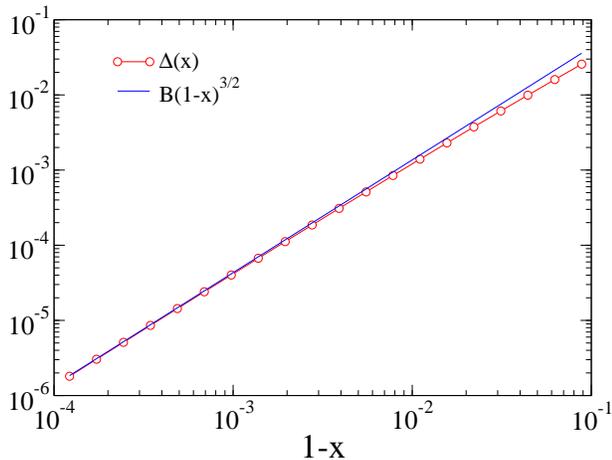}
\caption{The leading singular component of the generating function
${\cal C}_g(x)$ at the critical point. The quantity
\hbox{$\Delta(x)={\cal C}_g(x)-c_g+(1-x)$} and the fit $B(1-x)^{3/2}$
are plotted versus $1-x$.}
\label{fig-cg}
\end{figure}

Let $c_k$ be the density of clusters with total size $k$, and let
${\cal C}(x)=\sum_k c_k x^k$ be the corresponding generating
function. This single-variable generating function coincides with the
multivariate counterpart \eqref{cx-def} when all the coordinates are
identical
\begin{equation}
\label{diagonal}
{\cal C}(x)\equiv C(x,x,\ldots,x).
\end{equation}
Therefore, we have to vary $x$ and obtain $u\equiv u(x)$ by solving
the differential equation \eqref{u-eq} along with the integral
equation \eqref{root} with $x_i=x$ for all $i$. Finally, we substitute
$u$ into \eqref{cx-sol}.  We present results for the critical
behavior, ${\cal C}_g(x)\equiv {\cal C}(x,\tau_g)$, and focus on the
large-$k$ behavior, that follows from (figure \ref{fig-cg})
\begin{equation}
\label{cx-g}
{\cal C}_g(x)=c_g-(1-x)+B(1-x)^{3/2}+\cdots,
\end{equation}
as $x\to 1$.  The constant term is simply the critical concentration,
\hbox{$c_g=1-L_g=1-\langle j\rangle_g/2$} (Table I). The linear term
reflects the fact that finite clusters contain all the mass,
\hbox{${\cal C}'(1)=\sum_k k c_k=1$}. Finally, the leading singular
term $(1-x)^{3/2}$ shows that $c_k\sim k^{-5/2}$ as follows from the
Taylor series,
\begin{equation*}
(1-x)^{3/2}=\frac{3}{4\sqrt{\pi}}
\sum_{k=0}^\infty \frac{\Gamma(k-3/2)}{\Gamma(k+1)}\,x^k,
\end{equation*}
and the limiting behavior $\Gamma(k+a)/\Gamma(k+b)\simeq k^{a-b}$ when
$k\to\infty$. Hence, the critical cluster-size distribution has the
power-law tail
\begin{equation}
\label{ck-critical}
c_k\simeq A\,k^{-5/2},
\end{equation}
with $A=3B/4\sqrt{\pi}$ (see also Appendix A). Table II lists the
prefactors $B$ for $3\leq d\leq 7$ and $d=\infty$.  As a critical
phenomenon \cite{jes}, the evolving regular random graph is in the
universality class of mean-field percolation \cite{sa}.

\begin{table}[t]
\begin{tabular}{|c|c|c|c|c|c|c|}
\hline
$d$ & $3$ & $4$ & $5$ & $6$ & $7$ & $\infty$ \\
\hline
$B$ & $1.368$ & $1.037$ & $0.962$ & $0.944$ & $0.943$ & $0.942809$ \\
\hline
\end{tabular}
\caption{The prefactor $B$ in \eqref{cx-g} for $3\leq d\leq 7$ and
$d=\infty$. The last entry, $B=2\sqrt{2}/3$, is from
Eq.~\eqref{rg-cx-g}.}
\end{table}

\section{Large time behavior}
\label{LT}

We now turn to the large-time behavior. This behavior largely follows
from the degree distribution \eqref{nj-sol}.  We start with the crude
estimate, $\tau\simeq \ln t$, obtained by replacing the
right-hand-side of \eqref{tau-eq} with the dominant exponential term,
$d\tau/dt=e^{-\tau}$.  Then, we establish the corrections to this
logarithmic behavior by keeping the leading term in \eqref{tau-eq}
\begin{equation*}
\frac{d\tau}{dt}\simeq \frac{\tau^{d-1}}{(d-1)!}e^{-\tau}.
\end{equation*}
From this equation, we find the relation
\begin{equation*}
\frac{\tau^{d-1}}{(d-1)!}e^{-\tau}\simeq t^{-1},
\end{equation*}
when $t\to\infty$. Substituting this expression into \eqref{nj-sol}
gives the leading asymptotic behavior of the degree distribution
\begin{equation}
\label{nj-long}
n_j\simeq \frac{(d-1)!}{j!}\,t^{-1}\,(\ln t)^{-(d-1-j)},
\end{equation}
for $j<d$. The densities of active nodes with degree $j<d-1$ decay
algebraically with time, albeit with a logarithmic
correction. Furthermore, the degree distribution is an increasing
function of $j$. Since $\nu\simeq n_{d-1}$, the quantity $\nu$
exhibits the universal power-law decay
\begin{equation}
\label{nu-long}
\nu\simeq t^{-1}\,.
\end{equation}
The smaller the degree $j$, the faster the decay of $n_j$. In
particular, the density of isolated nodes exhibits the fastest decay, 
\begin{equation}
\label{n0-long}
n_0\simeq (d-1)!\,t^{-1}(\ln t)^{-(d-1)}.
\end{equation}
Each isolated node represents a minimal cluster with size
$k=1$. Therefore \eqref{n0-long} gives the density of minimal
clusters. The asymptotic behavior \eqref{n0-long} is much slower than
the fast exponential decay found in ordinary random graphs
\cite{krb}. Hence, the restriction on the degree leads to much slower
long-time kinetics.

The asymptotic results \eqref{nu-long} and \eqref{n0-long} provide
useful information about the final stages of the evolution. Amongst
all clusters, minimal clusters survive the longest. Hence, when all
minimal clusters disappear, a single cluster remains, and the giant
component takes over the entire system. The time to reach a single
component, $T$, grows with $N$ according to 
\begin{equation}
\label{T-sol}
T\sim \frac{N}{(\ln N)^{d-1}},
\end{equation}
as follows from the ``depletion'' criterion $Nn_0\sim 1$. This time is
not self-averaging: it fluctuates from realization to realization. In
particular, the average $\langle T\rangle $ does not give the second
moment $\langle T^2\rangle$; namely,
\begin{equation}
\label{self-averaging}
\lim_{N\to \infty}\frac{\langle T^2\rangle}{\langle T\rangle^2} >1.
\end{equation}
This breakdown of self averaging is in sharp contrast with the
behavior of ordinary random graphs where $T$ is a self-averaging
quantity.  The conclusion \eqref{self-averaging} is based on the
observation that a slight variation in the depletion criterion, say
$Nn_0\sim 2$, gives a different estimate for $T$.

\begin{figure}[t]
\includegraphics[width=0.45\textwidth]{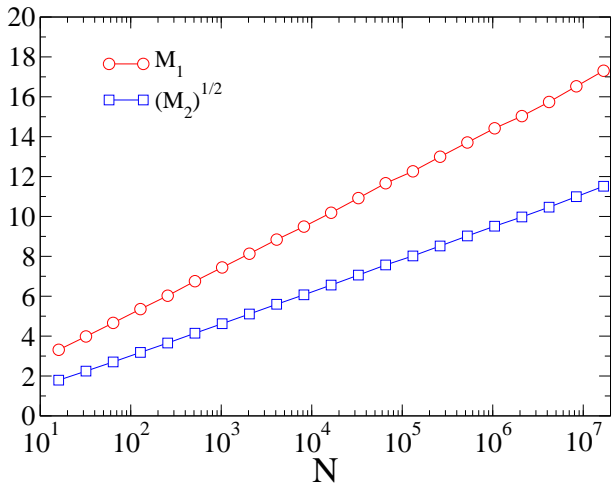}
\caption{The average number of active nodes when the graph becomes
fully connected for the first time. Shown are $M_1$ and $M_2^{1/2}$ as
a function of system size $N$, for the case $d=3$.  The results are
from an average over $10^4$ independent simulation runs.}
\label{fig-mj}
\end{figure}

When the giant component contains all $N$ nodes, the vast majority of
the nodes are inactive. Yet, a small majority of nodes remain
active. Let $M_j$ be the average number of nodes with degree $j>0$
when the graph becomes fully connected for the first time
\cite{sr-fp}. This quantity grows logarithmically with system size
(figure \ref{fig-mj})
\begin{equation}
\label{mj-sol}
M_j\sim (\ln N)^j,
\end{equation}
with $0<j<d$, as follows from $M_j\simeq Nn_j(T)$, along with
equations \eqref{nj-long} and \eqref{T-sol}. The majority of active
nodes have the largest possible degree, $j=d-1$.  As stated in
Eq.~\eqref{M-sol}, the total number of active nodes, when the giant
component first takes over the entire system, $M$, is logarithmic. 

To test these predictions, we performed Monte-Carlo simulations.  In
the simulations, we start with $N$ isolated nodes. At each step, we
pick two {\em active} nodes and connect them.  Keeping track of the
active nodes maximizes computational efficiency, and allows us to
simulate large systems with $N=10^7$.  We performed simulations for
the case $d=3$, and measured the average number of nodes with degree
$1$ and degree $2$ when the graph condenses into a single component
for the first time (figure \ref{fig-mj}). Also, we verified that the
time $T$ fluctuates from realization to realization by measuring the
ratio in \eqref{self-averaging}.

The asymptotic behavior \eqref{nj-long} indicates that there are
several intermediate steps until the regular random graph fully
forms. Isolated nodes disappear first, and at this time, there are
$(\ln N)^{d-1}$ active nodes. Nodes with degree $1$ disappear next,
and at this point, $(\ln N)^{d-2}$ active nodes remain. There are $d$
such steps, and in the last step, nodes with degree $d-1$
disappear. Now, the regular random graph is complete. The time it
takes to form the regular random graph is linear in system size,
\begin{equation}
t_f\sim N,
\end{equation}
as follows from the depletion criterion $N\nu \sim 1$, and the
algebraic decay \eqref{nu-long}. Like the quantity $T$, the formation
time $t_f$ is a fluctuating quantity, even in the limit
\hbox{$N\to\infty$}.

For completeness, we also mention the asymptotic behavior of the
cluster-size distribution,
\begin{equation}
\label{ck-long}
c_k\sim t^{-1} (\ln t)^{-k(d-1)},
\end{equation}
in the long-time limit ($d=2$ is a special case of a model considered
in \cite{mkr}). For minimal clusters, $c_1=n_0$, and this behavior
follows from the density of isolated nodes.  The behavior
\eqref{ck-long} follows from the master equation
\begin{equation}
\label{ck-eq-1}
\frac{dc_k}{dt} = \frac{1}{2}\sum_{l+m=k}lm\,c_lc_m - \nu kc_k\, ,
\end{equation}
for $k\leq d$.  Except for the loss rate, which is proportional to the
overall density of active nodes, $\nu$, this rate equation is the same
as \eqref{rg-ck-eq}. Clusters with $k\leq d$ contain only active
nodes, and hence, the loss term is proportional to cluster size,
$k$. Starting with $c_1\sim t^{-1}(\ln t)^{-(d-1)}$, the leading
asymptotic behavior \eqref{ck-long} can be established recursively
from equation \eqref{ck-eq-1}.

In general, we must classify clusters by their topology, and construct
specific rate equations for each topology.  The resulting rate
equations have the same overall structure as \eqref{ck-eq-1} and as a
consequence, the asymptotic behavior \eqref{ck-long} appears to hold
also when $k>d$. The proportionality constants depend on cluster
topology, however. Finally, we note that the criterion $Nc_k\sim 1$
shows that the asymptotic behavior \eqref{ck-long} is relevant only in
a limited range of sizes, $k\ll k_*$ with the logarithmic cutoff
$k_*\simeq (d-1)^{-1}\ln N$. The cluster size distribution is sharply
suppressed beyond the cutoff $k_*$.

\section{Discussion}
\label{disc}

In conclusion, we studied the evolution of random graphs with a
bounded degree using a dynamic linking process. First, a giant
component, which contains a macroscopic number of nodes, emerges in a
finite time. We obtained the nontrivial thresholds for this
percolation transition.  Then, the giant component takes over the
entire system. The evolution still proceeds as some nodes have yet to
reach the maximal degree. While the number of linking attempts before
the giant component emerges is a self-averaging quantity, the number
of steps it takes to form a fully connected graph or to complete the
regular random graph are both fluctuating (non-self-averaging)
quantities.

The regular random graph is completed via multiple steps.  In the
first step, nodes with degree $0$ disappear and at the point, the
graph is fully connected. Then, nodes with degree $1$ disappear,
etc. The total number of nodes with degree $d-1$ diminishes by a
factor $\ln N$ in each step,
\begin{equation*}
(\ln N)^{d-1}\to (\ln N)^{d-2}\to \cdots \to (\ln N) \to 0.
\end{equation*}
Similarly, the number of nodes of degree $d-2$ shrinks according to
$(\ln N)^{d-2}\to (\ln N)^{d-3}\to \cdots$. Furthermore, the asymptotic
behavior \eqref{ck-eq-1} indicates a multitude of time scales.
Clusters of size $k$ disappear at time
\begin{equation*}
T_k\sim \frac{N}{(\ln N)^{k(d-1)}},
\end{equation*}
for all $k\ll k_*$. The time $T\equiv T_1$ is simply the largest in a
series of time scales, $T_1>T_2>T_3>\cdots$. Therefore, multiple
finite-size scaling laws characterize kinetics of regular random
graphs.

The evolution equation for the cluster-size density is not a closed
equation. To address this challenge, we introduced the multivariate
size distribution, where the number of nodes with a given degree is
specified. For this quantity, the evolution equations are
closed. Using the Hamilton-Jacobi method, also useful for analyzing
large fluctuations in population dynamics \cite{fw,dmrh,ek,km}, we
constructed a formal solution for the multivariate size
distribution. Moreover, the cumbersome rate equations reduce to a
single second order differential equation, and numeric integration of
this equations gives the percolation thresholds with, essentially,
arbitrary accuracy.

Our analysis shows that the rate equation description extends to a
broader class of evolving random graphs.  Certainly, the multivariate
aggregation analysis can be generalized to situations where the
linking rate is degree dependent. The multivariate aggregation
framework can also be used to study the structure of clusters
\cite{bk04}, and in particular, cycles.

\smallskip
We thank Wolfgang Losert for useful discussions.  This research was
supported by DOE grant DE-AC52-06NA25396 and NSF grant CCF-0829541.

\appendix

\section{Classical Random Graphs}
\label{CRG}

This appendix briefly describes how to obtain relevant properties of
the classical random graph, $d=\infty$, using the Hamilton-Jacobi
formalism.  The density of connected components with size $k$ obeys
\cite{krb}
\begin{equation}
\label{rg-ck-eq}
\frac{dc_k}{dt} =\frac{1}{2}\sum_{l+m=k} lm\, c_l c_m -k c_k,
\end{equation}
with the initial condition $c_k(0)=\delta_{k,0}$.  The generating
function ${\cal C}(x,t)=\sum_k c_k(t) x^k$ satisfies
\begin{equation}
\label{rg-cx-eq}
\frac{\partial {\cal C}}{\partial t}+x\,\frac{\partial {\cal C}}{\partial
  x}=\frac{1}{2}\left(x\,\frac{\partial {\cal C}}{\partial x}\right)^2,
\end{equation}
with the initial condition ${\cal C}(x,0)=x$.

Using the momentum $p=\frac{\partial {\cal C}}{\partial x}$, the
Hamiltonian is
\begin{equation}
\label{rg-h-def}
H=xp-\frac{1}{2}(xp)^2.
\end{equation}
The Hamiltonian does not depend explicitly on time. Hence, $H$ is a
conserved quantity and consequently, $xp$ is also constant. The
evolution equations are
\begin{eqnarray}
\label{rg-xp-eq}
\frac{dx}{dt}=x(1-xp), \qquad \frac{dp}{dt}=-p(1-xp),
\end{eqnarray}
with the initial conditions $x(0)=y$ and $p(0)=1$. We can verify that
$d(xp)/dt=0$, and therefore, $xp=y$.  By integrating \eqref{rg-xp-eq},
we obtain the coordinate and the momentum,
\begin{eqnarray}
\label{rg-xp-sol}
x= y\,e^{(1-y)t},\qquad p= e^{-(1-y)t}.
\end{eqnarray}
The evolution equation for generating function is 
\begin{equation*}
\frac{d{\cal C}(x,t)}{dt} =- H + p \frac{\partial H}{\partial p}=
-\frac{1}{2}(xp)^2.
\end{equation*}
Hence, the generating function is a quadratic function of the initial
coordinate,
\begin{equation}
\label{rg-cx-sol}
{\cal C}(x,t)=y-\frac{1}{2}y^2t,
\end{equation}
with $x=y\,e^{(1-y)t}$.

The mass of the giant component, $g=1-\sum_k k c_k$, follows from the
momentum $p$ at $x=1$.  The conservation law $xp=y$ implies that $p=y$
if $x=1$.  In analogy with \eqref{g-def}, we have $g=1-y$, and using
\eqref{rg-xp-eq}, we find a closed equation,
\begin{equation}
\label{rg-g-sol}
g=1-e^{-gt}.
\end{equation}
There is a trivial solution $g=0$, valid at all time, and a second
nontrivial solution, $g>0$, valid only when $t>1$. The latter is
physical when $t>1$, and hence, the giant component emerges at time
$t_g=1$.

To obtain the critical size distribution, that is, the size density at
$t=t_g$, we note that the critical generating function is given by
\hbox{${\cal C}_g(x)=y-\frac{1}{2}y^2$} with $x=ye^{1-y}$.  We then
focus on the behavior for $x\to 1$. We substitute the expansion
\hbox{$1-y=a_1(1-x)^{1/2}+a_2(1-y)+\cdots$} into the equation
$x=y\,e^{1-y}$, and solve for the coefficients $a_1$ and $a_2$. This
calculation gives
\begin{equation*}
1-y=[2(1-x)]^{1/2}-\frac{2}{3}(1-x)+\cdots.
\end{equation*}
This expansion, together with Eq.~\eqref{rg-cx-sol}, gives the leading
behavior of the critical generating function,
\begin{equation}
\label{rg-cx-g}
{\cal C}_g(x)=\frac{1}{2}-(1-x)+\frac{2\sqrt{2}}{3}(1-x)^{3/2}+\cdots,
\end{equation}
in the limit $x\to 1$. Thus, the critical size distribution has the
power-law tail \cite{zhe} 
\begin{equation}
c_k\simeq \frac{1}{\sqrt{2\pi}}k^{-5/2},
\end{equation}
as follows from the leading singular component $(1-x)^{3/2}$.


\begin{thebibliography}{99}
\bibitem{tree}   More complicated clusters do occur, but they are very
                 rare, e.g. in classical random graphs their total
                 number scales logarithmically with system size, see
                 e.g. \cite{jlr,bk05}.
\bibitem{smooth} The density $c(t)$ is twice differentiable at the
                 percolation point; the third derivative undergoes a
                 jump \cite{krb}.
\end{thebibliography}
\end{document}